# Algebraic order and the Berezinskii-Kosterlitz-Thouless transition in an exciton-polariton gas

Wolfgang H. Nitsche,[1,*] Na Young Kim,[1] Georgios Roumpos,[1,†] Christian Schneider,[2] Martin Kamp,[2] Sven Höfling,[2,3,4] Alfred Forchel,[2] and Yoshihisa Yamamoto[1,4,‡]

[1]*E. L. Ginzton Laboratory, Stanford University, Stanford, California 94305, USA*
[2]*Technische Physik, Universität Würzburg, Am Hubland, 97074 Würzburg, Germany*
[3]*School of Physics & Astronomy, University of St Andrews, St Andrews KY16 9SS, United Kingdom*
[4]*National Institute of Informatics, Hitotsubashi, Chiyoda-ku, Tokyo 101-8430, Japan*

We observe quasi-long range coherence in a two-dimensional condensate of exciton-polaritons. Our measurements are the first to confirm that the spatial correlation algebraically decays with a slow power-law, whose exponent quantitatively behaves as predicted by the Berezinskii-Kosterlitz-Thouless theory. The exciton-polaritons are created by non-resonant optical pumping of a micro-cavity sample with embedded GaAs quantum-wells at liquid helium temperature. Michelson interference is used to measure the coherence of the photons emitted by decaying exciton-polaritons.

Bose-Einstein condensation is characterized by the macroscopic occupation of the lowest-energy state and has been observed in atomic gases [1,2] as well as with quasi-particles in solid state systems [3,4]. It is accompanied by superfluidity [5-8] and long-range spatial coherence [9,10]. According to the Hohenberg-Mermin-Wagner theorem, true long-range order cannot exist in infinite uniform two-dimensional (2D) superfluids at non-zero temperatures [11,12]. However in interacting systems, superfluidity can persist at non-zero temperature, in a state exhibiting quasi-long-range order characterized by an algebraic (power-law) decay of the spatial correlation function. The transition from this state to the normal state is described by the Berezinskii-Kosterlitz-Thouless (BKT) theory [13,14], and is explained by the creation of free vortices which eliminate both the superfluidity and any quasi-long-range coherence [15]. In finite sized systems, the quasi-long-range order can span the whole system, resembling true long-range order. Nevertheless, the transition from this state to the normal one is still expected to be of the BKT type. Previously, this transition has been demonstrated for superfluid liquid helium films [16,17], superconducting films [18,19] and 2D atomic gases [20]. However, the power-law decay of spatial correlations expected for the BKT phase, which is its most distinct characteristic, has never been established quantitatively. In this letter, we show that the measured first-order spatial correlation function of an exciton-polariton condensate decays with a power-law, whose exponent is $\approx 1/4$ at the condensation threshold, as predicted by the BKT theory. Recent theoretical work [21] confirmed that the continuous creation and decay of exciton-polaritons does not prevent our finite-sized non-equilibrium system from exhibiting the BKT state with quasi-long-range coherence.

The first order spatial correlation function

$$g^{(1)}(\mathbf{r}_1, \mathbf{r}_2) = \frac{\langle \psi^\dagger(\mathbf{r}_1)\psi(\mathbf{r}_2) \rangle}{\sqrt{\langle \psi^\dagger(\mathbf{r}_1)\psi(\mathbf{r}_1) \rangle \langle \psi^\dagger(\mathbf{r}_2)\psi(\mathbf{r}_2) \rangle}} \qquad \text{(equation 1)}$$

quantifies the coherence between points $\mathbf{r}_1$ and $\mathbf{r}_2$. Its values range from 0 in the case of no coherence to 1 for perfect coherence. In the case of 2D systems, coherence can be reduced by the thermal excitation of phononic long wavelength phase fluctuations as well as vortices.

Vortices are characterized by a vanishing superfluid density in their cores and a phase change of $2\pi$ (or $-2\pi$ in the case of an antivortex) around them. Large 2D superfluids always contain vortices, the presence of which decreases the free energy by increasing the entropy. These vortices can appear either in the form of bound vortex-antivortex pairs or as free vortices. The BKT theory predicts that the existence of free vortices is advantageous with respect to the free energy precisely if the 2D superfluid density $n_s$ is less than the critical value of $4/\lambda_T^2$, where $\lambda_T = h/\sqrt{2\pi m_{\text{eff}} k_B T}$ is the thermal de Broglie wavelength [13,14]. This means that vortex-antivortex pairs unbind once the superfluid density drops below the critical value [15], either by decreasing $n_s$ or increasing the temperature. Thus created free vortices completely destroy the spatial coherence and the superfluid phase [13,14], so that $n_s$ must be either exactly zero or larger than the critical value. Therefore a 2D superfluid with non-vanishing spatial coherence can only exist if $n_s$ is above the critical value. In this case, the superfluid is in the BKT state, which means that vortices only exist in the form of bound vortex-antivortex pairs. These pairs do not affect the spatial coherence over large distances because the phase disturbance of each vortex is cancelled out by its counterpart of antivortex.

The coherence over long distances in the BKT state is determined solely by the residual thermal excitation of phononic long-wavelength phase fluctuations. These phonons are predicted [15,22] to cause the coherence $g^{(1)}$ between $\mathbf{r}_1 = (x; y)$ and $\mathbf{r}_2 = (-x; y)$ to decay as a power-law of the form

$$g^{(1)}(x,-x) = \frac{n_s}{n}\left(\frac{|x|}{\Lambda}\right)^{-a_p} \qquad \text{(equation 2)}$$

where $n_s$ and $n$ are the superfluid and total densities, $\Lambda$ is a characteristic length of the order of the healing length $\xi$ and the exponent is $a_p = 1/(n_s \lambda_T^2)$. This power-law decay of $g^{(1)}$ is specifically a result of the density of states for phonons in a 2D system [15], and it differs distinctly from the three-dimensional BEC state where the correlation function approaches a constant value at large distances [23]. The critical density of $n_s = 4/\lambda_T^2$ implies that $a_p = 1/4$ at the BKT threshold and $a_p < 1/4$ in the BKT state above threshold, whereas below threshold no exponent $a_p$ can be defined due to the absence of any long range order. For our measurements, we use an exciton-polariton condensate, for which it is possible to study the coherence properties for varying superfluid densities not only near the critical value but also far above that.

Exciton-polaritons are 2D bosonic quasi-particles which can be described as the superposition of cavity photons and quantum well excitons [24]. At low temperatures, they exhibit dynamical condensation [3,4,25]. This condensate of exciton-polaritons can show a vortex-antivortex bound pair [26] as well as single vortices pinned at a defect [27-30]. Exciton-polaritons decay if a photon leaks out of the sample; in this case the leaking photons preserve the energy, in-plane momentum and coherence properties of the polaritons and the internal order parameter. Michelson interference allows us the measurement of $g^{(1)}$ of the leaking photons from the exciton-polariton condensate (Fig. 1), which makes an exciton-polariton condensate a suitable system for studying the BKT physics.

The sample used in our measurements is grown on a GaAs wafer and consists of a $\lambda/2$ AlAs cavity which is surrounded by Bragg reflectors with 20 AlAs/AlGaAs layer pairs at the top of the sample and 24 such pairs between the cavity and the substrate. Four 7 nm-thick GaAs quantum wells, separated from each other by 4 nm AlAs spacers, have been grown into the central antinode of the standing photon wave in the cavity. The $\lambda/2$ cavity has been grown with a wedge form so that the detuning parameter $\delta = (\omega_{cavity} - \omega_{exciton})\hbar$ changes with position. The splitting between the upper and lower polaritons at zero detuning is approximately 9.6 meV and the quality factor of the cavity mode (as measured in the far red detuned regime on the planar sample) is $Q_{cavity} \approx 3000$. The sample is installed in a helium flow cryostat which keeps the temperature at 5 K.

We create an exciton-polariton condensate (Supplementary Fig. S1 and S2 [31]) by non-resonantly pumping the sample with perpendicular incidence by a Ti:Sapphire laser in continuous-wave (CW) operation. The pump wavelength is shorter than the cavity resonant wavelength, and is chosen to coincide with a reflection minimum outside the Bragg reflector stop band so that the pump light can efficiently reach the quantum wells where it creates electron-hole pairs. To avoid thermal heating, the laser is chopped at 100 Hz with a duty-cycle of 5% so that during each $10\,\mu s$ period the pump laser only hits the sample for $0.5\,\mu s$, which is still much longer than the exciton-polariton lifetime, so that CW excitation is guaranteed.

For the measurements described in this letter, we use a laser beam with a Gaussian spatial intensity profile (with a full width at half maximum of $15\,\mu m$) which can only efficiently excite the lowest spatial mode of the condensate. Previous measurements [22] used a different sample, and tried to create a uniform condensate density by using a top-hat formed pump beam. However such a top-hat pump profile can easily lead to condensate fragmentation into multiple spatio-energy modes [32]. In this case, the measured overall visibility might decay faster than the intrinsic coherence for each individual mode, since any interference between different energy modes averages out to zero during the integration time of the camera.

To measure the coherence of the emitted photons, we use a Michelson interferometer (Fig. 1(a)) with a variable path-length difference $L$ as described in Ref. [22]. A polarizing beam-splitter is used to direct the pump beam onto the sample through an objective lens. The same objective lens collects the light emitted by the sample, and the component which is polarized perpendicular to the pump beam passes through the polarizing beam-splitter and reaches the Michelson interference setup. There it is divided by a non-polarizing beam splitter into the two arms of the interferometer. The light which travels through the first arm is reflected by a mirror and directed towards the camera, whereas the light in the other arm is reflected by a reflection prism which flips the image along the $y$-axis before directing it to the camera. Therefore at the camera the light which is emitted at point $(x;y)$ of the sample and travels through the first arm interferes with the light from point $(-x;y)$ coming through the second arm. The wave-fronts from the two arms reach the camera at a slightly different angle which gives rise to interference fringes. Scattered light from the pump laser is filtered out by band-pass filters before reaching the camera. A piezo is used for changing the path-length difference by slightly moving the prism, and the interferogram is recorded for many different path-length differences. The measured intensity of each of the pixels (corresponding to a position $(\pm x; y)$ on the sample) as a function of the change $\Delta L$ of the path-length difference $L$ follows a sine law of the form $I_{\text{measured}}(\Delta L) = B + A\sin(w\Delta L - \varphi_0)$. As explained in the first chapter of the supplementary information [31], the first order correlation function $g^{(1)}(x,-x)$ is identical to the visibility $V(x)$ and can therefore be calculated as $g^{(1)}(x,-x) = A/B$ from the fitting parameters $A$ and $B$. The same sine fit also gives the phase $\varphi_0$ (Fig. 2(a)) which can be used to confirm that the long-range-order is produced and the fit was able to extract reliable data of the visibility [22].

A typical result (Fig. 2(a-c)) for the measured phase and fringe visibility shows three distinct regions: The visibility over short distances (I) decays according to a Gaussian law. This is similar to the correlation function decay of a Bose gas in thermal equilibrium [23] where the width of the Gaussian is proportional to the de Broglie wavelength. However in our system the population of energetically higher modes depends primary on their overlap with the Gaussian pump profile, therefore we cannot deduce an accurate temperature from this Gaussian width [22]. In the region of intermediate distances (II), the decay of the visibility follows a power law as theoretically predicted for the BKT state. From this, we determine the exponent $a_p$ by fitting a power-law proportional to $|x|^{-a_p}$ to the measured fringe visibility in this region. For even larger distances (III), the visibility starts to decay much faster, which we attribute to the decrease of the superfluid fraction towards the edge of the condensate. The intensity of the Gaussian pump decreases towards the edge, which is expected to lead to a decrease of the superfluid fraction $n_s/n$. According to equation 2 (and as shown in Supplementary Fig. S5 [31]), this picture can explain the fast drop of $g^{(1)}$.

A pump-power dependent measurement of the fringe visibility shows that, as predicted by the BKT theory, the exponent $a_p$ is approximately $1/4$ at threshold and decreases with increasing pump-power (Fig. 3). This figure also shows that overall the exponents $a_p$ extracted for both $x > 0$ and $x < 0$ are equivalent, even if each individual measurement of the visibility (as shown in Fig. 2(c)) might appear slightly asymmetrical due to experimental imperfections. By recording the position resolved image (Supplementary Fig. S1 [31]) of the condensate with a calibrated camera and estimating the transmittance $T_{\text{path}}$ of the beam-path from the sample to the camera, we can estimate the photon flux $j$ at which photons leak out of the sample. The exciton-polariton lifetime $\tau_{\text{exciton-polariton}}$ can be calculated as $\tau_{\text{exciton-polariton}} \approx \tau_{\text{cavity}}/|C|^2$, where the Hopfield-coefficient $|C|^2$ is the detuning dependent photon component of the exciton-polariton, and the photon lifetime $\tau_{\text{cavity}}$ is known from the quality factor of the cavity. The 2D total exciton-polariton density is estimated as $n = j\tau_{\text{exciton-polariton}}$. We know that above the threshold, the superfluid density $n_s$ is about one half of this total density $n$, since the measured visibility is approximately $1/2$ at the transition from region I to region II in Fig. 2(c). By fitting a thermal distribution to the measured energy resolved population (Supplementary Fig. S3 [31]), although the system is not in thermal equilibrium, we can estimated an effective temperature of the exciton-polaritons to be always $\approx 25\,\text{K}$, from which we get the thermal de Broglie wavelength $\lambda_T$. The continuous black line in Fig. 3 shows the calculated value for $a_p^{\text{calculated}} := 1/(n_s \lambda_T^2)$ using these experimental values of $n_s$ and $\lambda_T$, and gives a reasonable match with the experimental values for $a_p$ from the interference measurements.

In Fig. 4(b), we show one-to-one correspondence between the measured exponents $a_p$ and the phase-space density of $1/(n_s \lambda_T^2)$, as well as the predicted continuous line with $a_p^{\text{calculated}} = 1/(n_s \lambda_T^2)$. The exponents $a_p$, which have been measured for different pump-powers and detuning parameters, follow the predicted universal line. This observation indicates that the 2D exciton polariton condensate behaves as expected from the BKT theory.

Our measurements confirm that exciton-polariton condensation can be understood not by the BEC model but exclusively within the BKT model. The measured first-order spatial coherence function fully supports a unique 2D superfluid characteristic both at and above threshold.

The authors acknowledge helpful discussion with E. Altman, J. Dalibard, M. Fraser, J. Keeling and W. Phillips. This research has been supported by the Japan Society for the Promotion of Science (JSPS) through its "Funding Program for World-Leading Innovative R&D on Science and Technology (FIRST Program)", by Navy/SPAWAR Grant N66001-09-1-2024, and by National Science Foundation ECCS-09 25549. W. H. N. acknowledges the Gerhard Casper Stanford Graduate Fellowship.

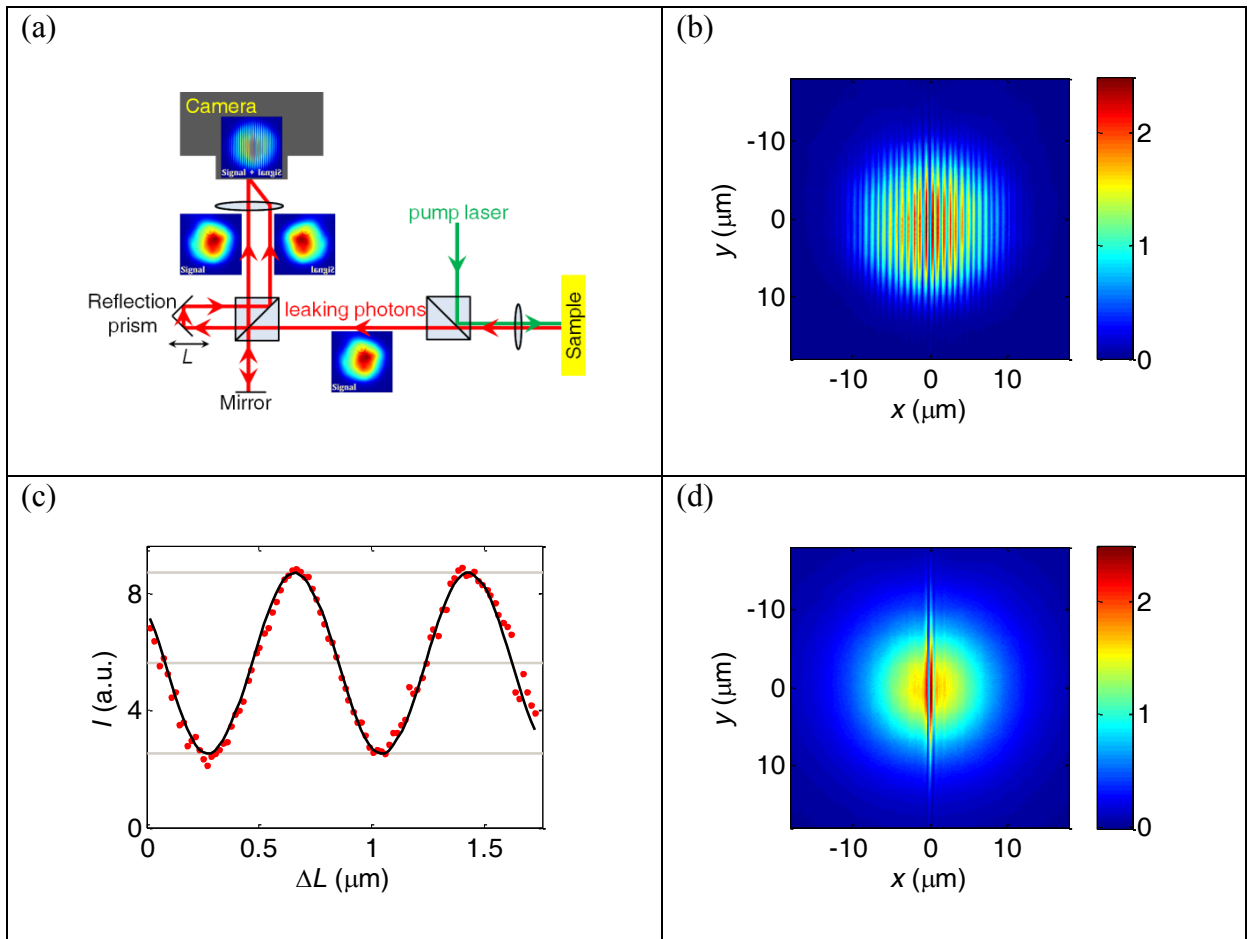

FIG. 1. Michelson interference.
(a) Michelson interferometer setup to measure the phase and the fringe visibility
(b) Measured interferogram (with intensity in arbitrary units) for one specific path-length difference $L$, measured at a pump-power above condensation threshold. (Same measurement as left column of Fig. 2)
(c) The intensity of each pixel behaves like a sine function of the form $I_{\text{measured}}(\Delta L) = B + A\sin(w\Delta L - \varphi_0)$ if plotted as a function of the change $\Delta L$ of the path-length difference $L$.
(d) Same as (b), but at a low pump-power below condensation threshold where no spatial coherence exists over long distances.

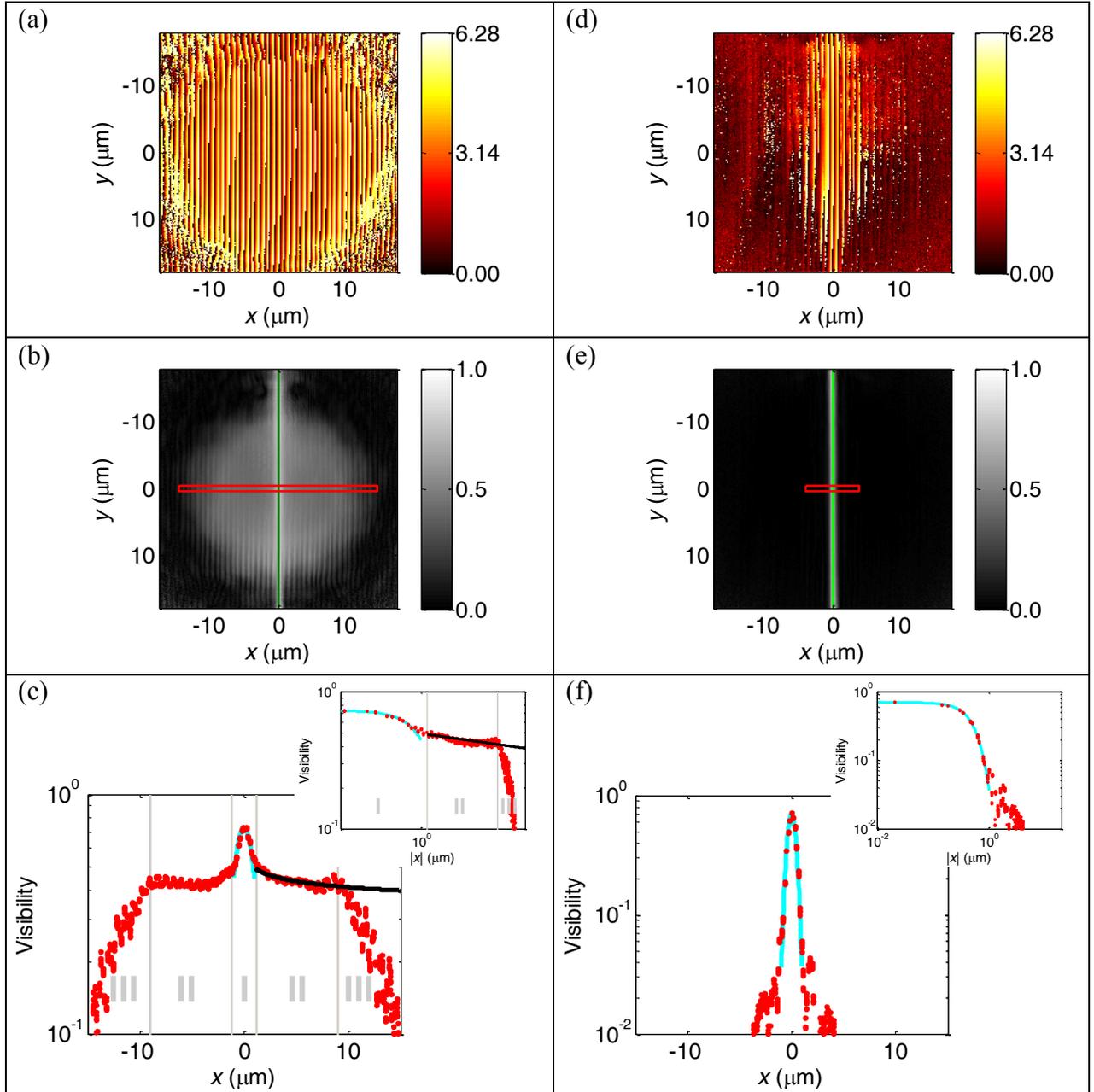

FIG. 2. Measured phase and visibility (measured at detuning parameter $\delta \approx -1\,\text{meV}$). Left column: above condensation threshold (at $p_{\text{pump}} = 14\,\text{meV}$)

(a) Phase map $\varphi_0$ produced by an ensemble of pixels.

(b) Visibility map. The flip-axis is shown as a line at $x=0$ and the region of interest used for further evaluation is highlighted by a rectangle.

(c) The visibility within the region of interest as a function of the directed distance $x$ to the flip-axis. In region II, a power-law fit with exponent $a_p = 0.082$ is shown. The regions I, II, and III are selected manually. (Insert: The same in a log-log plot where the power-law decay appears as a straight line.)

Right column: the same characteristics at below condensation threshold (at $p_{\text{pump}} = 2\,\text{mW}$).

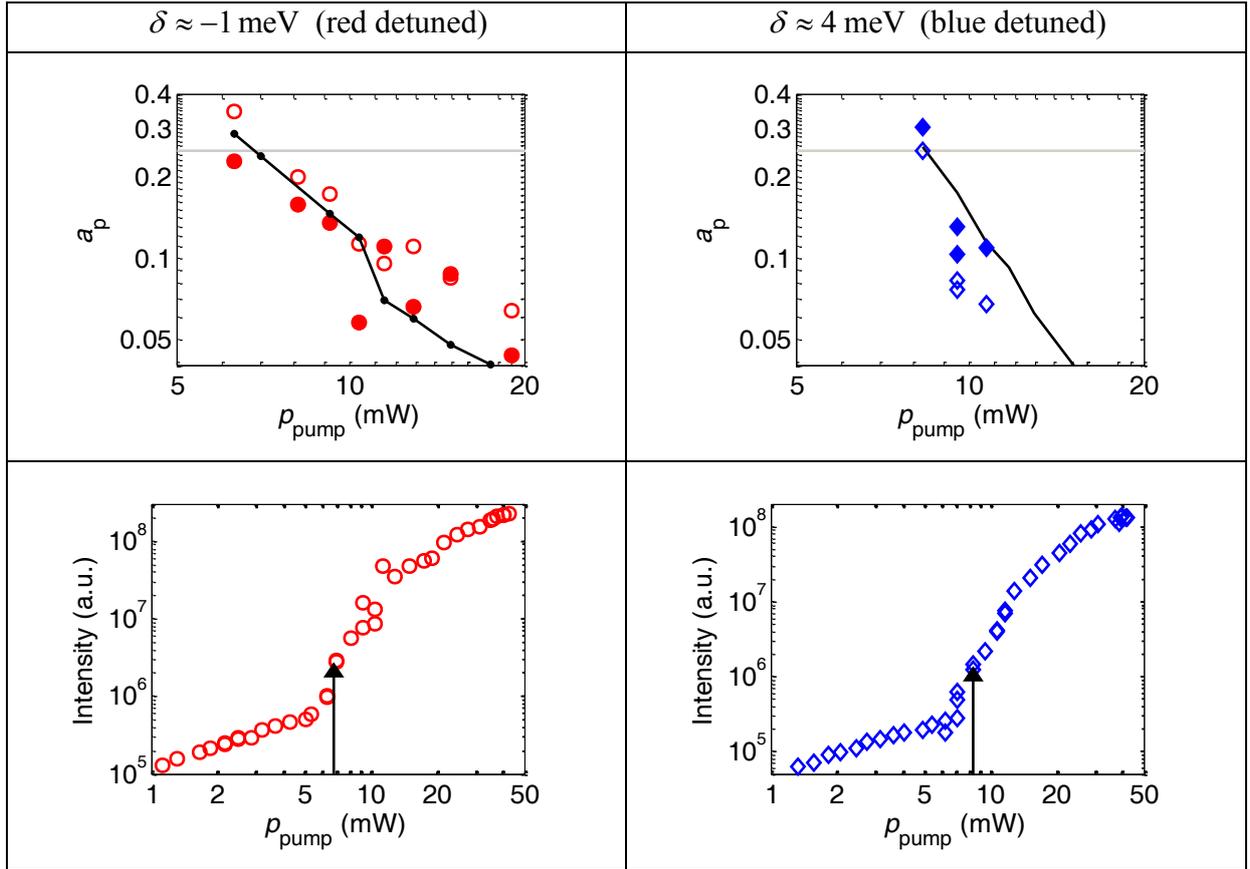

FIG. 3. Pump-power dependence.
Upper row: Pump-power dependence of the exponent $a_p$ for two different detuning values. The symbols show the measured exponents as determined by a power-law fit as shown in Fig. 2(c). The grey horizontal line at 0.25 shows the theoretically predicted exponent at threshold. The black continuous line shows the estimated inverse superfluid phase-space-density $1/(n_s \lambda_T^2)$, and as predicted by the theory, it matches with the measured exponents $a_p$. Filled symbols correspond to the $x<0$ region.
Lower row: Pump-power dependent peak intensity, as determined by a energy and momentum resolved measurement. The threshold pump-powers defined by the critical exponent $a_p = 0.25$ (upper traces) are indicated by the black arrows.

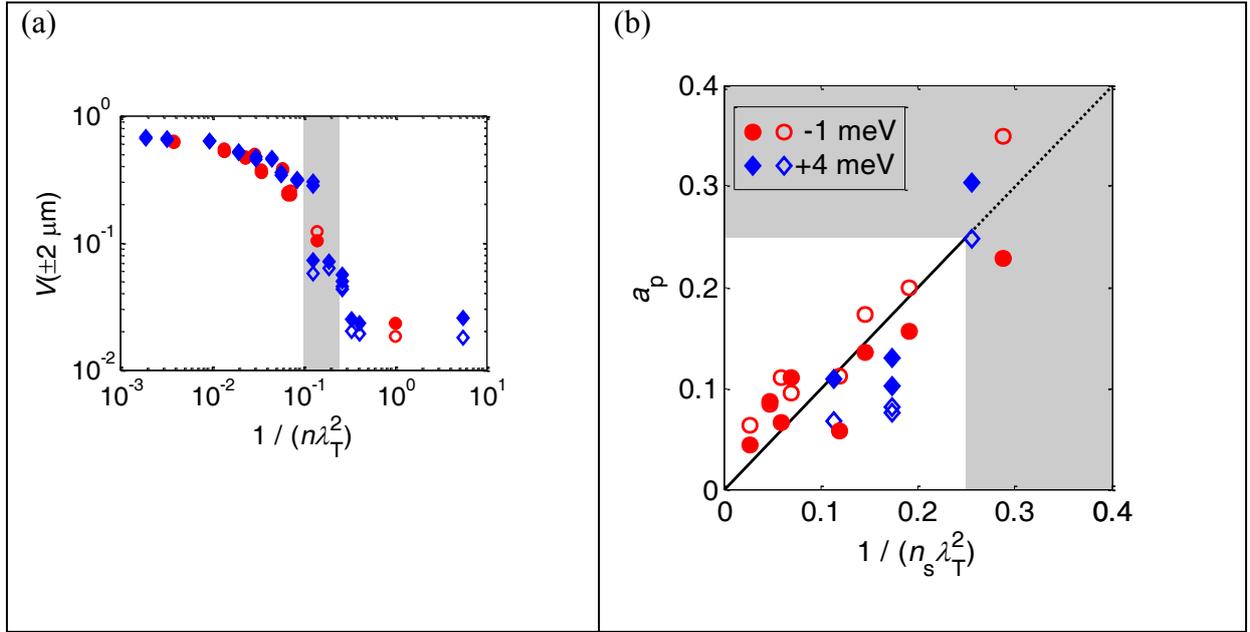

FIG. 4. Phase space density, visibility and power-law exponent.
As in Fig. 3, filled symbols correspond to the $x<0$ region, and different kinds of symbols to different detunings.

(a) The visibility at $x = \pm 2$ μm vs. the inverse phase-space-density $1/(n\lambda_T^2)$. The threshold is upper bounded by $1/(n_s\lambda_T^2) = 1/4$ and lower bounded by $1/\ln\left[380\hbar^2/(m_{\text{eff}} g_{\text{interaction}})\right] = 1/10$, which are shown by the shaded area.

(b) Measured exponents $a_p$ as a function of estimated inverse superfluid phase-space-density $1/(n_s\lambda_T^2)$. The black line shows the predicted $a_p^{\text{calculated}} = 1/(n_s\lambda_T^2)$, and the region above the expected threshold of 0.25 is shaded in gray. Although we did not change the temperature during the measurement, the abscissa can be interpreted as the dimensionless temperature (in units of $2\pi m_{\text{eff}} k_B n_s^{-1} \hbar^{-2}$, and the BKT phase transition occurs at the dimensionless temperature of 0.25 in these units.

# Supplemental material for:
# Algebraic order and the Berezinskii-Kosterlitz-Thouless transition in an exciton-polariton gas

Wolfgang H. Nitsche,[1] Na Young Kim,[1] Georgios Roumpos,[1] Christian Schneider,[2] Martin Kamp,[2] Sven Höfling,[2,3,4] Alfred Forchel,[2] and Yoshihisa Yamamoto[1,4]

[1]E. L. Ginzton Laboratory, Stanford University, Stanford, California 94305, USA
[2]Technische Physik, Universität Würzburg, Am Hubland, 97074 Würzburg, Germany
[3]School of Physics & Astronomy, University of St Andrews, St Andrews KY16 9SS, United Kingdom
[4]National Institute of Informatics, Hitotsubashi, Chiyoda-ku, Tokyo 101-8430, Japan

**Visibility and $g^{(1)}$; Irregular condensate for high pump-powers**

The measured visibility (as shown in Fig. 2(b)) corresponding to $(x;y)$ can be calculated as

$$V(x;y) = \frac{2\sqrt{I_1(x;y)I_2(-x;y)}}{I_1(x;y)+I_2(-x;y)} g^{(1)}(x,y,-x;y) \quad \text{(equation S1.1)}$$

where $I_1(x;y)$ is the intensity of the signal which reaches the camera after being emitted from point $(x;y)$ and travelling through arm 1 of the interferometer and likewise $I_2$ travels through arm 2.

With $I_1(x;y) = f_1 I(x;y)$ where $I(x;y)$ is the total emitted intensity at $(x;y)$ and $f_1$ is the fraction travelling through arm 1 (and likewise for arm 2), the visibility becomes

$$V(x) = \frac{2\sqrt{f_1 I(x) f_2 I(-x)}}{f_1 I(x) + f_2 I(-x)} g^{(1)}(x,-x) \quad \text{(equation S1.2)}$$

where we omitted the constant $y$.

In theory $f_1$ and $f_2$ should be identical, but in reality this usually not the case, for example because the mirror in arm 1 has a different reflectivity than the reflection-prism in arm 2. In our setup, we have $f_2 \approx 0.80 f_1$ which, for $I(x) = I(-x)$, as expected for a symmetrical condensate, gives $V(x) = 0.994 g^{(1)}(x,-x)$, meaning that the differences between $f_1$ and $f_2$ can be neglected.

Assuming $f_1 \approx f_2$ gives

$$V(x) = \frac{2\sqrt{I(x)I(-x)}}{I(x)+I(-x)} g^{(1)}(x,-x) \quad \text{(equation S1.3)}$$

which means that the visibility $V$ will decay faster than $g^{(1)}$ if $I(x) \neq I(-x)$ due to an irregular form of the condensate. For a perfectly symmetrical condensate with $I(x) = I(-x)$, we get

$$V(x) = g^{(1)}(x, -x) \quad \text{(equation S1.4)}$$

which means that under optimal conditions the measured visibility $V$ is identical to the correlation function $g^{(1)}$.

As shown in Fig. 3, for pump-powers moderately above condensation threshold, increasing the pump-power decreases the exponent $a_p$ of the power law. For even higher pump-powers, the measured $a_p$ seems to increase, instead of showing the expected steady decrease. This increase can be explained as an artefact resulting from the spatial density of the condensate, which at high pump-powers deviates from a Gaussian profile and becomes increasingly non-symmetric (Fig. S1) so that $I(x) \neq I(-x)$, which causes the measured visibility to decay faster than $g^{(1)}$. Therefore, for Fig. 3 (upper row) and Fig. 4(b), we did not consider data from very high pump-powers corresponding to an irregular condensate form.

We believe that the irregularities of the condensate form are caused by local fluctuations of the sample properties. In the case of irregular condensate forms, we see certain patterns like dark lines going through the condensate, and if we slightly move the sample, these patterns move with the sample, which indicates that they are caused by local properties of the sample.

In theory, the visibility is always expected to be symmetrical, so that $V(x) = V(-x)$. However equation S1.2 shows that under realistic experimental conditions, $V(x) \neq V(-x)$ is possible if $I(x) \neq I(-x)$ and at the same time $f_1 \neq f_2$. This can explain the slightly asymmetric data in Fig. 2(c).

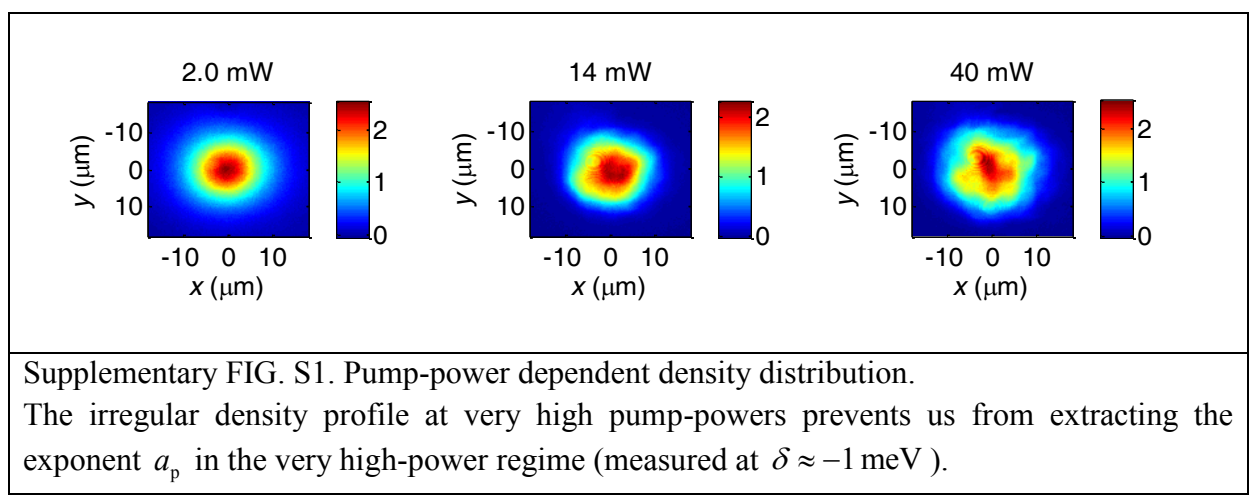

Supplementary FIG. S1. Pump-power dependent density distribution.
The irregular density profile at very high pump-powers prevents us from extracting the exponent $a_p$ in the very high-power regime (measured at $\delta \approx -1\,\text{meV}$).

**Power-dependent blue-shift of exciton-polariton energy**

In Fig. S2(a), one sees that the exciton-polaritons in the condensate are more energetic than those measured below condensation threshold. Fig. S2(b) shows the power-dependent signal which is emitted close to $k=0$. A jump to shorter wavelengths at condensation threshold as well as a continuous blue-shift above threshold are the result of repulsive interaction between exciton-polaritons. The continuous blue-shift above threshold is a signature of exciton-polariton condensation since it does not appear in the case of VCSEL lasing.

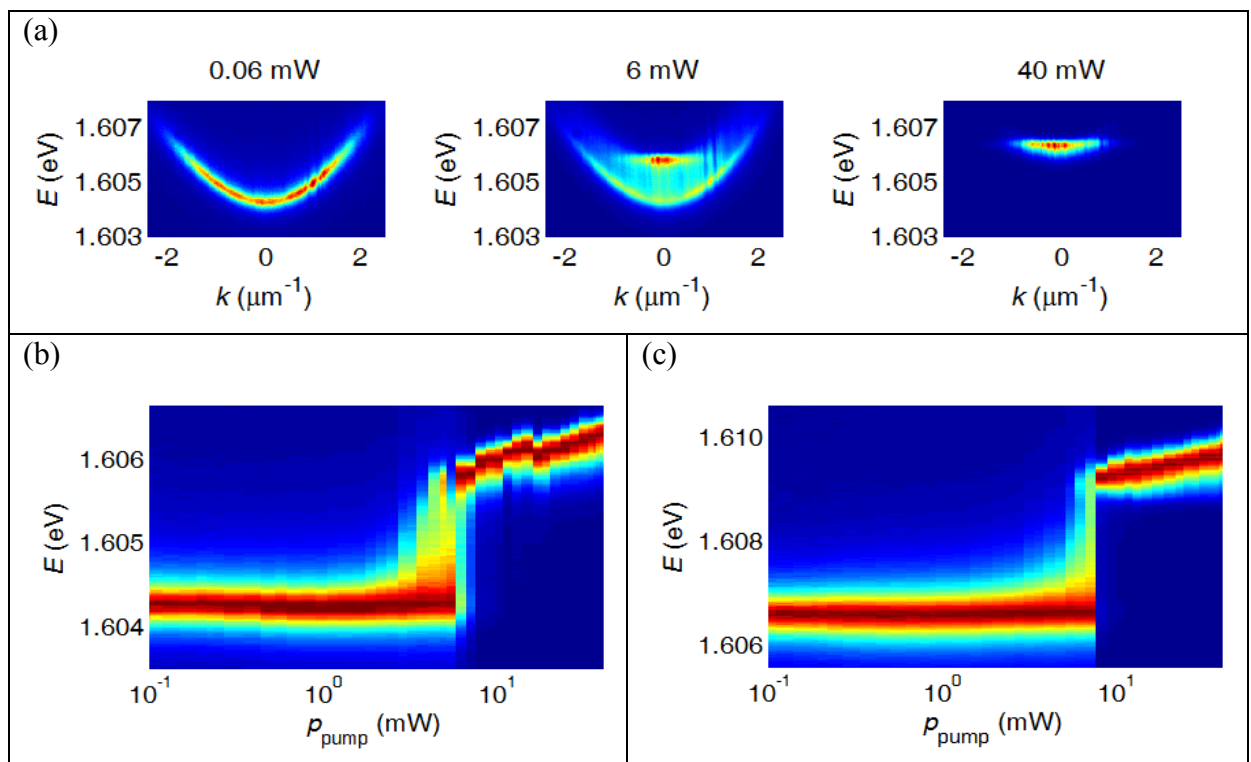

Supplementary FIG. S2. Blue-shift of exciton-polariton energy.
(a) Pump-power dependent dispersion (measured at $\delta \approx -1\,\text{meV}$). At a threshold of 6 mW, the condensation phase appears, and at above this pump-power, nearly all exciton polaritons are in a ground state with zero momentum and one specific energy. A blue-shift towards higher energies due to repulsive interaction between the exciton-polaritons is observed above condensation threshold.
(b) Normalized signal close to zero momentum at blue detuning of $\delta \approx -1\,\text{meV}$. The measurement of the exciton-polariton density, which we described in the main part of this letter, indicates that at the threshold of $p_{\text{pump}} \approx 6\,\text{mW}$, the total exciton polariton density is $n \approx 2\,\mu\text{m}^{-2}$, which corresponds to an exciton-polariton density per quantum-well of $n_{\text{per QW}} \approx 0.5\,\mu\text{m}^{-2}$. This small value confirms that we observe condensation rather than VCSEL lasing, since the later requires significant higher exciton-polariton densities of approximately $3\times 10^3\,\mu\text{m}^{-2}$ per quantum-well to create a population-inversion.
(c) The same at red detuning of $\delta \approx 4\,\text{meV}$.

**Determination of the exciton-polariton temperature**

While the sample is kept at 5 K, exciton-polaritons, which are not at thermal equilibrium with the lattice, have a different temperature. Their temperature can be estimated from a position $x$ and energy $E$ resolved measurement, by fitting a Bose-Einstein distribution to the measured occupation (Fig. S3). From this, we estimate their temperature to be $T \approx 25$ K.

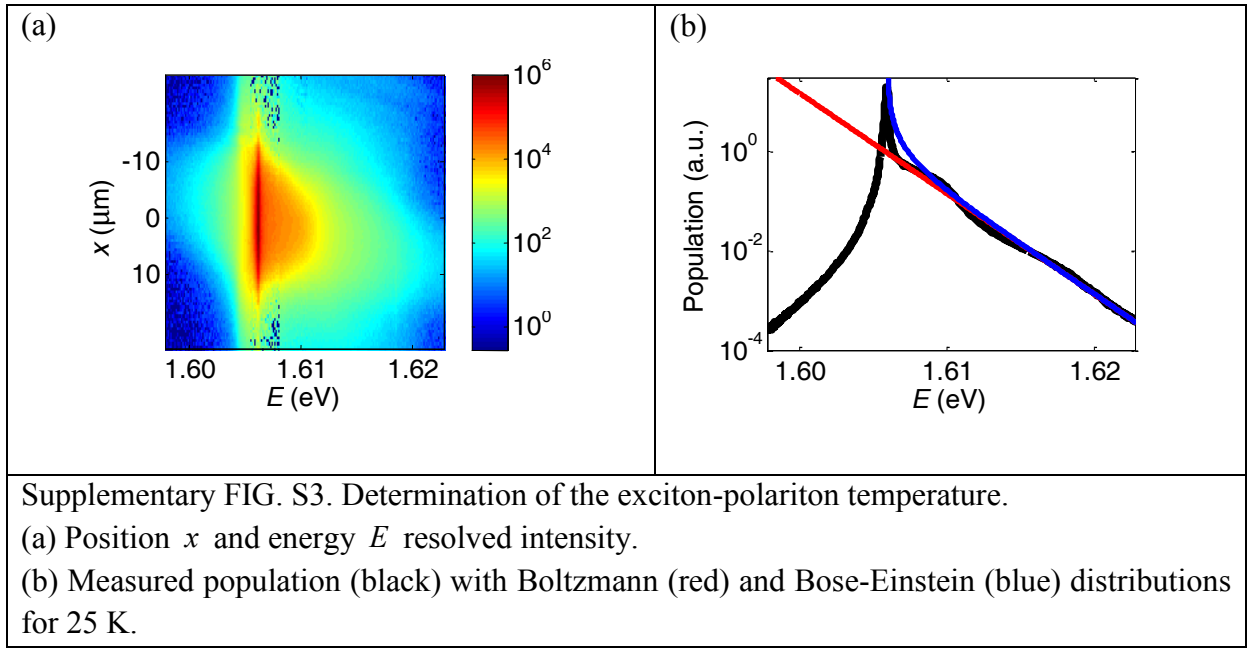

Supplementary FIG. S3. Determination of the exciton-polariton temperature.
(a) Position $x$ and energy $E$ resolved intensity.
(b) Measured population (black) with Boltzmann (red) and Bose-Einstein (blue) distributions for 25 K.

## Transition from power-law to faster decay

In Fig. 2(c), we see that the power-law (II) ends at a "cut-off"-length beyond which a much faster decay (III) is observed. This data has been extracted close to $y = 0$ in Fig. 2(b). By performing the same evaluation for different $y$-values, we see that the "cut-off"-points seem to lie on a circle (Fig. S4), which confirms our interpretation that the fast decay of the visibility is caused by a decrease of the condensation-fraction towards the edge of the condensate.

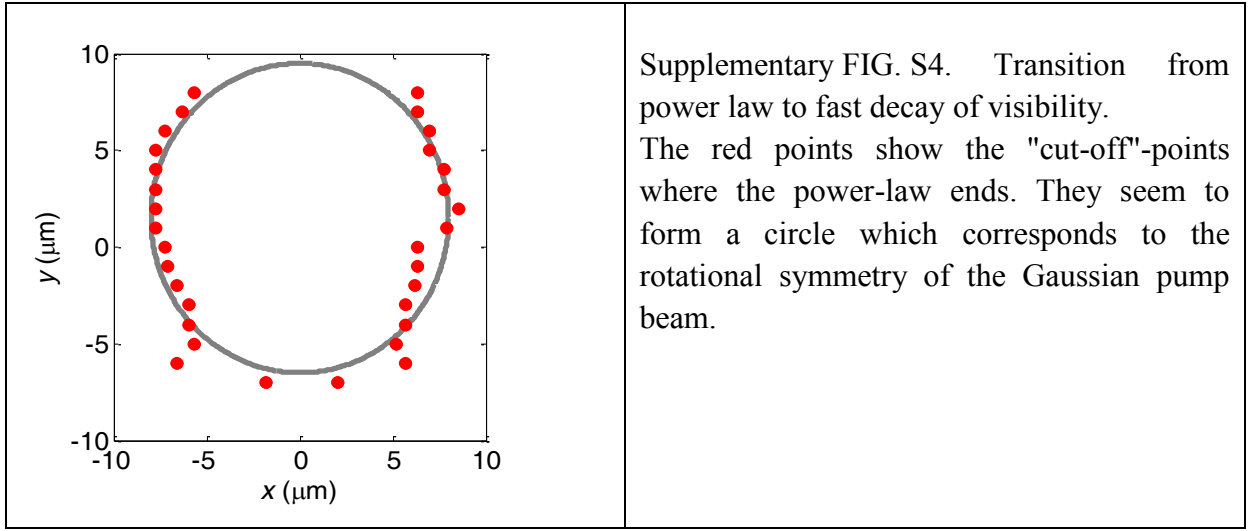

Supplementary FIG. S4. Transition from power law to fast decay of visibility.
The red points show the "cut-off"-points where the power-law ends. They seem to form a circle which corresponds to the rotational symmetry of the Gaussian pump beam.

## Simulated visibility

To demonstrate that the fast decay in region III of Fig. 2(c) can be explained by a decreasing superfluid fraction $n_s/n$ at the edge of the condensate, we simulated the visibility (Fig. S5). For this, we assumed a condensate with a radius of $15\,\mu\text{m}$, where the superfluid fraction is constant for $r < 10\,\mu\text{m}$ and slowly decreases to a small value at at $r = 15\,\mu\text{m}$, where it jumps to 0%.

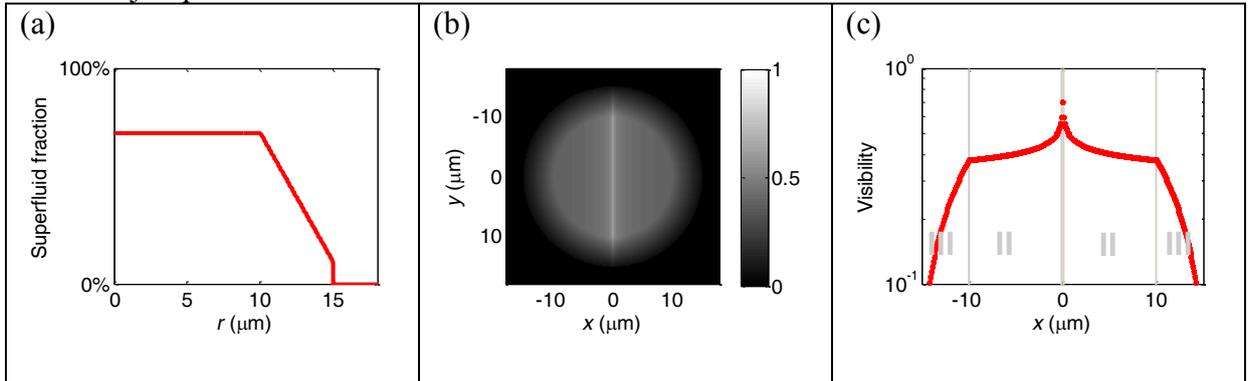

Supplementary FIG. S5. Simulated visibility.
(a) Assumed position dependent superfluid fraction as a function of radial position $r$.
(b) Simulated visibility assuming $g^{(1)} = (n_s/n)(|x|/0.02\,\mu\text{m})^{-0.1}$.
(c) Simulated visibility at $y \approx 0$. The simulated fast decay in region III shows that the fast decay in Fig. 2(c) can be explained by a decreasing superfluid fraction.

## Simulated power-law

The theory for the BKT phase predicts that the power-law decay of the correlation function $g^{(1)}$ is the result of thermally excited phonon modes whereas all vortices are paired so that they cannot affect $g^{(1)}$. We simulated a two-dimensional condensate whose phase fluctuations are caused by phonons with wavelengths of at least $2\pi$ times the healing length (Fig. S6). For this simulation, we assumed that the number of phonons in the different possible modes follow a phonon statistic, meaning that the probability to have exactly $N_\mathbf{k}$ phonons in the mode $\mathbf{k}$ is

$$p_{N_\mathbf{k}} = \frac{\exp\left(-\frac{N_\mathbf{k}\varepsilon_\mathbf{k}}{k_\mathrm{B}T}\right)}{1-\exp\left(-\frac{\varepsilon_\mathbf{k}}{k_\mathrm{B}T}\right)}$$

where $\varepsilon_\mathbf{k} = \hbar c |\mathbf{k}|$ is the energy per phonon in this mode $\mathbf{k}$. The local phase can be expressed [113] in the form $\Theta(\mathbf{r}) = \alpha_\mathbf{k} \sin(\mathbf{k}\mathbf{r} + \varphi_\mathbf{k})$ where the Fourier amplitudes $\alpha_\mathbf{k}$ depend on the numbers $N_\mathbf{k}$ of phonons in the respective modes. No vortices have been considered in this simulation. As expected, the simulated visibility decays with a power-law, which confirms that this characteristic decay can be caused by thermally excited phonons, rather than by the presence of vortices. As shown in Ref. [113], the same result can also be derived analytically.

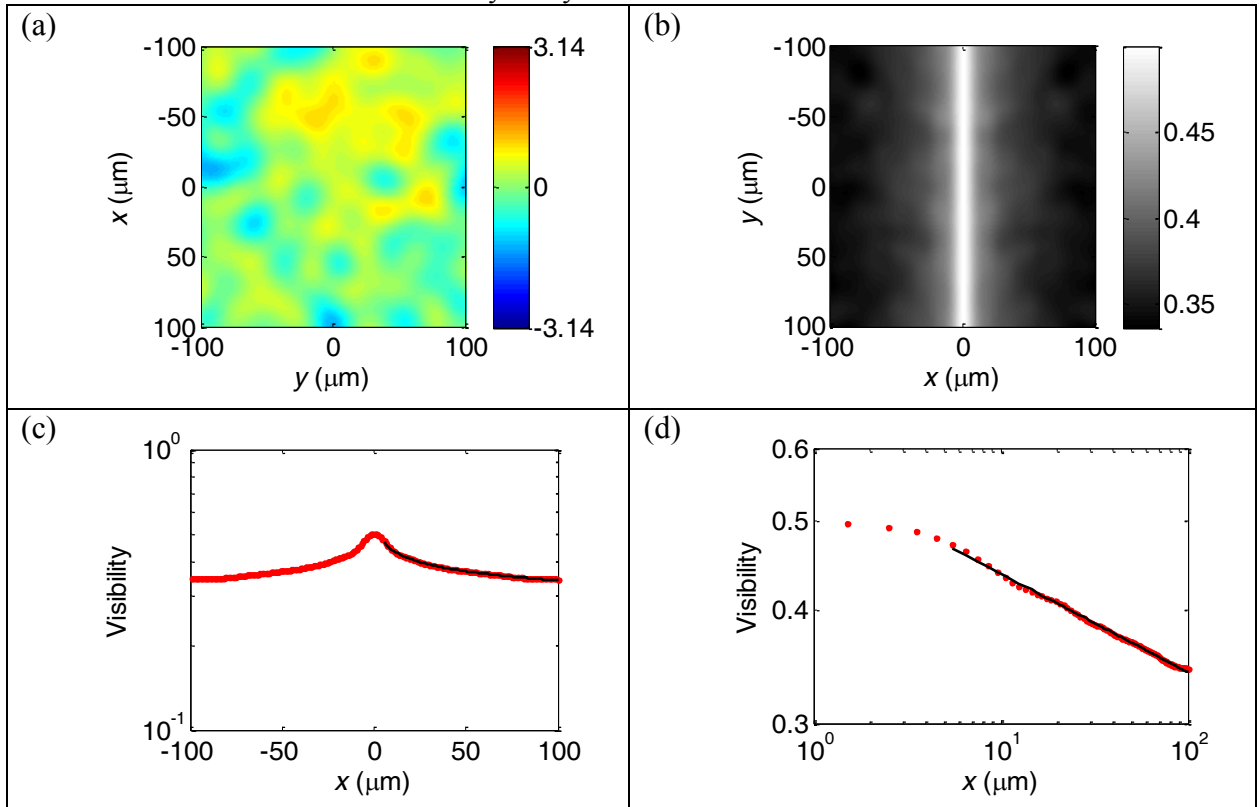

Supplementary FIG. S6. Simulated power-law.
(a) Phase $\Theta$ in one run of the simulation.
(b, c) Visibility averaged over 1000 runs of the simulation.
(d) The same visibility shown as a log-log plot where the power-law appears as a straight line.